\documentclass[fleqn,usenatbib]{mnras}
\pdfoutput=1
\usepackage{newtxtext,newtxmath}

\usepackage[T1]{fontenc}
\usepackage{ae,aecompl}

\usepackage{graphicx}	
\usepackage{amsmath}	
\usepackage{amssymb}	
\usepackage[dvipsnames]{xcolor}

\title[Vertical gradient of radial migration]{Radial migration and vertical action in $N$-body simulations}

\author[D. Mikkola, P. J. McMillan, D. Hobbs]{
Daniel Mikkola\thanks{E-mail: mikkola@astro.lu.se },
Paul J. McMillan, 
David Hobbs
\\
Lund Observatory, Lund University, Department of Astronomy and Theoretical Physics, Box 43, SE-22100, Lund, Sweden\\
}

\date{Accepted XXX. Received YYY; in original form ZZZ}

\pubyear{2019}

\begin{document}
\label{firstpage}
\pagerange{\pageref{firstpage}--\pageref{lastpage}}
\maketitle

\begin{abstract}
We study the radial migration of stars as a function of orbital action as well as the structural properties of a large suite of $N$-body simulations of isolated disc galaxies. Our goal is to establish a relationship between the radial migration efficiency of stars and their vertical action. We aim to describe how that relationship depends on the relative gravitational dominance between the disc and the dark matter halo.
By changing the mass ratio of our disc and dark matter halo we find a relationship between disc dominance, number and strength of spiral arms, and the ensuing radial migration as a function of the vertical action. We conclude that the importance of migration at large vertical action depends on the strength of the spiral arms and therefore the dominance of the disc. Populations with more radial action undergo less radial migration, independently of disc dominance. Our results are important for the future of analytical modelling of radial migration in galaxies and furthers the understanding of radial migration which is a key component of the restructuring of galaxies, including the Milky Way.

\end{abstract}




\section{Introduction}\label{sec:intro}
During the evolution of a galaxy there are a number of external and internal factors which play a part in shaping its chemodynamical structure. One of these factors is called radial migration and is capable of displacing stars over large radial distances. Because of this radial migration plays an important part in the restucturing of a galaxy over time. In this paper we will study radial migration over large timescales to determine which stars migrate and in what way this is affected by the strength and number of the spiral arms present.

A galaxy can suffer mergers with other galaxies and has significant evolution from within, through Giant Molecular Clouds (GMCs) and secular features like bars and spiral arms. Both in the context of an isolated galaxy and when there is a dynamic galactic environment the local regions of a galaxy do not evolve independently. This has been clearly seen in studies of the chemical properties of the Milky Way, particularly in the age-metallicity relationship  \citep{edvardsson,bensby14,bergemann}. Studies of this nature show that there is a significant scatter in abundances at almost all ages. Such a scatter would not exist in an isolated setting and instead supports the existence of a restructuring process.

An important process that restructures galaxies is radial migration. \citet{sellwood02} showed that significant angular momentum changes could occur when the pattern speeds of stars and spirals match at co-rotation, in addition to the angular momentum changes at the Lindblad resonances previously known \citep{lynden}. This process is able to move stars by kiloparsecs and does not leave dynamical traces.

The importance of radial migration has been shown not only by its role in broadening abundance distributions across the Galaxy, but has been suggested as an explanation for the bimodality of stars in the [$\alpha$/Fe]-[Fe/H] plane \citep{schön09,toyouchi} and as cause for mono-age population flaring in the outer disc \citep{minchev2012,minchev2015}. The effects of radial migration has been studied extensively by simulations \citep{roskar2008,halle2015,aumer1,aumer2,aumer3,aumer4}, analytical models \citep{sellwood02,schön09,schön&mcmill}, and real data \citep{minchev2018,neige} but still requires further understanding of which stars are more likely to undergo migration.


Radial migration is driven by secular features such as spiral arms and bars and is therefore going to affect stars differently depending on their positions and velocities within the disc of the galaxy. Studies that utilize analytical models like \citet{schön09} and \citet{schön&mcmill} therefore rely upon a sound understanding of which stars are migrated and to which extent. The extent of radial migration as a function of position or velocity about the midplane has been studied previously by, e.g. \citet{solway}, \citet{vera-ciro14,vera-ciro16} and in the paper \citet{kathryne} the link between radial migration and dynamical temperature was investigated. 



In this paper, we perform a large suite of $N$-body galaxy simulations to probe radial migration in terms of kinematics and its effects on the structure of galactic discs. We look at the migration of stars as a function of their vertical and radial actions, $J_z$ and $J_r$, which quantifies the oscillations about the midplane of the disc and the average radius along an orbit respectively. The structure of this paper is as follows; in Section \ref{sec:theory} we outline the two processes commonly referred to as radial migration, in Section \ref{sec:sims} we go through the details of the simulations we have performed and a subsequent Fourier analysis of them, in Section \ref{sec:results} we present our simulation results in terms of structural properties, actions, and action conservation before comparing them to those of \cite{solway} and \cite{vera-ciro14,vera-ciro16} who studied closely related topics, and in Section \ref{sec:conc} we give our conclusions.

\section{Radial migration}\label{sec:theory}
We will consider the two processes most commonly referred to as radial migration in disc galaxies namely \textit{blurring} \citep{schön09} and \textit{churning} \citep{sellwood02}. Both are processes related to the orbits of stars are significantly different. In this section we approximate for simplicity's sake that the Galaxy is a flat disc and use angular momentum to refer to the vector perpendicular to this disc, $L_z = Rv_\phi$, where $R$ and $v_\phi$ are the radius and azimuthal velocity of a particle respectively.

\textit{Blurring} is the change in amplitude of radial oscillations around an average radius for an orbit, called the guiding radius, $R_g$. A star will be born on a nearly circular orbit and very likely scattered at some point in its life from a GMC or similar, placing it onto a slightly more or less radially extended orbit. The guiding radius and therefore angular momentum, since $L_z$ is directly relatd to $R_g$, does not change through this process, and it is the change in amplitude of the oscillations between closest and furthest galactic radius that define blurring. 

The second source of radial migration is \textit{churning}, first described by \citet{sellwood02}, which is caused by non-axisymmetric features such as bars and spiral arms exerting a torque on a star. In contrast to blurring, churning can change the angular momentum of an orbit without changing its eccentricity. \cite{sellwood02} showed that conservation of the Jacobi integral, $E_J=E-\Omega_{\rm p}L_z$, in the presence of a steady non-axisymmetric perturbation with pattern speed $\Omega_{\rm p}$ implies the relationship
\begin{equation}
\Delta J_R = \frac{\Omega_p - \Omega}{\omega_R}\Delta L_z.
\end{equation}
Here $\Delta J_R$ is a measure of the extent of radial oscillations. $\Delta L_z$ is the change in angular momentum, $\Omega_p$ is the pattern speed of the spiral/bar, $\Omega$ is the angular speed of a star, and $\omega_R$ is the frequency of radial oscillations. A change in $\Delta J_R$ will follow from radial migration where the angular momentum is changed, $\Delta L_z \neq 0$, but is made less pronounced if such a migration occurs near co-rotation where the angular velocity of the star is the same as the spiral arm/bar, $\Omega = \Omega_p$, in which case there is close to zero change in radial action. This feature of churning is perhaps also the most frustrating, as it means stars are able to radially migrate with no dynamical trace of the procedure, which removes the possibility of dynamically discerning a migrated star in the Solar neighbourhood from a local one. The direction of the migration is determined by whether the star is inside or outside the co-rotation resonance as the sign of the exerted torque will change. A star inside co-rotation moves outwards and vice versa. A more rigorous demonstration of churning was given in \cite{sellwood02} with spiral arms churning stars and gas without changing the overall angular momentum distribution or increasing the random motions significantly.

We mentioned in the previous section the importance of understanding which stars migrate and this is not fully understood. It is important to further increase our understanding of which gradients in radial migration exist if we are to use analytical models of it. In \citet{kathryne} they use a series of analytical disc galaxy models and study the fraction of stars trapped at co-rotation depending on the velocity dispersion.  They find that this fraction declines with increasing radial velocity dispersion. Their analysis is 2D and therefore does not make any statements about the vertical distribution of migrators. 

Two articles that study the vertical gradient of radial migration are \cite{solway} and \cite{vera-ciro14}. \cite{solway} uses an $N$-body disc in a static potential halo. They include both a thick and thin disc and conclude that the root mean square angular momentum changes are gradually reduced. That radial migration is reduced by vertical motion is also supported by \citet{vera-ciro14} and subsequently by \citet{vera-ciro16} who use three different simulations of live discs embedded in static halo potentials. These systems vary in the dominance of the disc and this creates different spiral morphologies \citep[e.g. ][]{dongia}. The result in all three simulations is what is called a `provenance bias', i.e, that migration primarily concerns stars with small vertical excursions regardless of spiral pattern. This is at first glance not surprising since stars with small vertical excursions should spend more time closer to the midplane where the spirals are strongest. However, sufficiently strong spirals, which could arise in very disc dominated systems, could migrate stars of larger vertical excursions as well. How much radial migration is a function of its vertical excursions depending on the spiral morphology and disc dominance has not previously been established. To investigate this, we use a large suite of simulations that span a broad range in disc dominance. We use $N$-body discs as well as halos and bulges and investigate the effects on spiral strength and structure. To probe the effect of vertical excursions on radial migration we calculate actions for our stars and compare the vertical action, $J_z$, to the amount of migration that occurs in different parts of our galaxies. 

\subsection{Action variables}\label{sec:theory-actions}
Instead of characterising the vertical and radial gradients of radial migration by position or velocities we use actions. Take for example the vertical action:
\begin{equation}
J_z = \frac{1}{2\pi}\oint_{\gamma_z} v_z \cdot z,
\end{equation}
where the line integral is over a path, $\gamma_z$, of phase-space coordinates that the orbit 
can go through and which goes through a full vertical oscillation. This can be, for example, the heights and vertical velocities ($z$,  $v_z$) that an orbit can have as it goes through a given radius $R$ \citep[the surface of section, e.g., ][\S 3.2.2]{galdyn}. However, it is enormously difficult to calculate the actions using a surface of section directly. We use a more accurate and robust approach available in the software library \textsc{agama}. This allows us to approximate the gravitation potential of our simulation as an axisymmetric expansion in spherical harmonics, and to calculate actions by applying the ``St\"{a}ckel fudge'' from \cite{staeckel}, which uses the approximation that any orbit in a realistic galactic potential can be closely approximated by one in a St\"{a}ckel potential.

This allows us to use vertical action instead of vertical position and velocity which has the advantage of not oscillating along an orbit and is a good measure of how vertically heated an orbit is. The same is true for $J_r$ in the radial direction.

\section{Simulations}\label{sec:sims}
In order to study radial migration we generate a number of $N$-body galaxies which were numerically integrated using the tree code \textsc{gyrfalcON} \citep{gyrfalcon1,gyrfalcon2} which is available as part of the \textsc{nemo}\footnote{\label{nemo}\url{https://github.com/teuben/nemo}} \citep{nemo} toolbox.

\subsection{Initial conditions}
\label{sec:maths}
To generate the initial conditions, three packages within \textsc{nemo} were used: \textsc{mkWD99disc}, \textsc{mkhalo}, and \textsc{mkgalaxy} \citep{mkwd99disc}, the first two for the disc and halo/bulge respectively and the latter combines the two for systems containing a disc, halo, and bulge.
\subsection*{The disc}
The disc is generated using the procedure laid out in \cite{mkwd99disc}. This method uses the distribution function described in \citet{WD99} which has the advantage that it avoids using a Maxwellian approximation, and therefore starts close to equilibrium. The disc is generated iteratively in the potential of the halo and bulge, tending towards the specified profile. It is designed to be stable to axisymmetric perturbations, but not stable to the non-axisymmetric perturbations that are expected to arise.
The density profile is of the form
\begin{equation}
\rho_\mathrm{disc}(R,z) = \frac{1}{2z_d}\Sigma_0 \exp \left(-\frac{R}{R_d}\right) \mathrm{sech}^2\left(\frac{z}{z_d}\right), 
\end{equation} 
where the disc mass corresponds to $M_d = 2\pi R_d^2 \Sigma_0$ with $\Sigma_0$ being the scale density, $R_d$ is the scale radius, and $z_d$ is the scale height. The radial velocity dispersion profile is $\sigma_R \propto \exp (-R/R_\sigma)$ where the parameters $R_\sigma$ and $Q$, the selected value for Toomre's Q \citep{toomre}, determine the constant of proportionality. The vertical velocity dispersion is $\sigma_z^2 = \pi G\Sigma(R)z_d$. The number of bodies in a single orbit is set with $N_\mathrm{bpo}$ and the gravitational softening length is $\epsilon$. The parameters used when generating the disc are listed in table \ref{tab:disc_params}. 
\begin{table}
	\centering
	\caption{Parameters of the $N$-body disc. The mass of the disc, $M_d$, the number of particles, $N_d$, the scale length, $R_d$, the scale height, $z_d$, the normalization constant of $\sigma_R$, Toomre's $Q$, the number of bodies per orbit, and the softening length, $\epsilon$.}
	\label{tab:disc_params}
	\begin{tabular}{cccccccc} 
		\hline
		$M_d$ & $N_d$ & $R_d$ & $z_d$ & $R_\sigma$ & $Q$ & $N_\mathrm{bpo}$ & $\epsilon$\\
		
		[M$_{\odot}$] &  & [kpc] & [kpc] & [kpc] &  &  & [kpc]\\
		\hline
		$5\times 10^{10}$ & $10^6$ & 3 & 0.3 & 9 & 1.7 & 50 & 0.03\\
		\hline
	\end{tabular}
\end{table}
\begin{table}
	\centering
	\caption{Parameters of the fiducial dark matter halo and bulge components. The mass of the component, $M_\mathrm{tot}$, the scale radius, $r_s$, the truncation radius, $r_t$, the total number of particles in the component, $N_\mathrm{tot}$, the inner exponent, $\gamma_i$, the outer exponent, $\gamma_o$, the transition exponent, $\eta$, and  the softening length, $\epsilon$.}
	\label{tab:sph_params}
	\begin{tabular}{cccccccc} 
	    \hline
        $M_\mathrm{tot}$  & $r_s$ & $r_t$ & $N_\mathrm{tot}$ & $\gamma_i$ & $\gamma_o$ & $\eta$ & $\epsilon$ \\
        
        [M$_{\odot}$] & [kpc]  & [kpc] &  &  &  &  & [kpc]\\ \hline
        \multicolumn{8}{c}{\textbf{Halo}}     \\
        $3.4\times 10^{11}$ & 17 & 30 & $2\times 10^6$ & 7/9 & 31/9 & 4/9 & 0.02  \\ \hline
        \multicolumn{8}{c}{\textbf{Bulge}}    \\
        $1.5\times 10^{10}$ & 1 & 3 & $5\times 10^5$ & 1 & 4 & 1 & 0.03 \\
		\hline
	\end{tabular}
\end{table}
\subsection*{The halo and bulge}
Both halo and bulge are generated within \textsc{mkgalaxy} and the spheroids are created with spherical density distribution:
\begin{equation}
    \rho(r) = \frac{\rho_0}{x^{\gamma_i}(x^\eta + 1)^{\gamma_o - \gamma_i / \eta}}\mathrm{sech}\left(\frac{r}{r_t}\right),
\end{equation}
with $x=r/r_s$, $r_s$ being scale radius, $\rho_0$ is the scale density, $r_t$ is the truncation radius, $\gamma_i$ and $\gamma_o$ are the inner and outer exponents, and $\eta$ is the transition strength between them. For the halo these parameters are selected to produce a Dehnen-McLaughlin dark matter halo \citep{DMhalo} which can be seen in table \ref{tab:sph_params}. Further explanation of the parameters can be found in \citet{mkwd99disc}. The sech factor of the profile is a truncation. For the bulge a Hernquist profile \citep{hernquist} is used with parameters in table \ref{tab:sph_params}. To constrain the parameters for the different components and achieve Milky Way-like initial conditions, values from \citet{2017mcmillan} are used. The mass of the halo is altered from this fiducial galaxy setup, as discussed below.

\subsection{Simulation types}
\begin{figure}
	\includegraphics[width=0.48\textwidth]{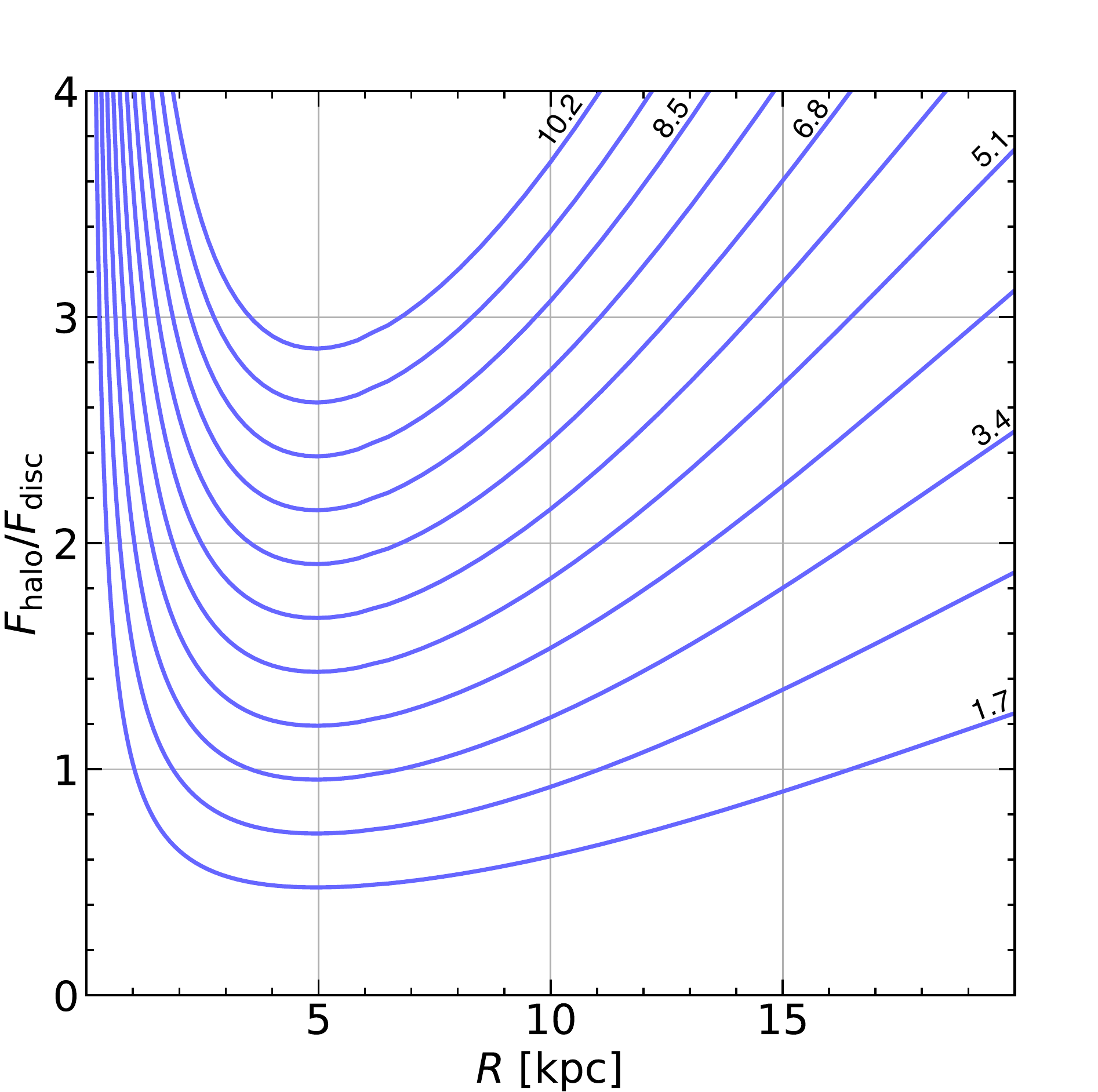}
    \caption{Ratio of the radial force from the dark matter halo and the disc at $z=0$ against radius in simulations varying in dark matter halo total mass. The mass of the dark matter halo is written above its respective line for every other simulation in units of $10^{11} M_\odot$.}
    \label{fig:frate}
\end{figure}
\begin{figure*}
	\includegraphics[width=1\textwidth]{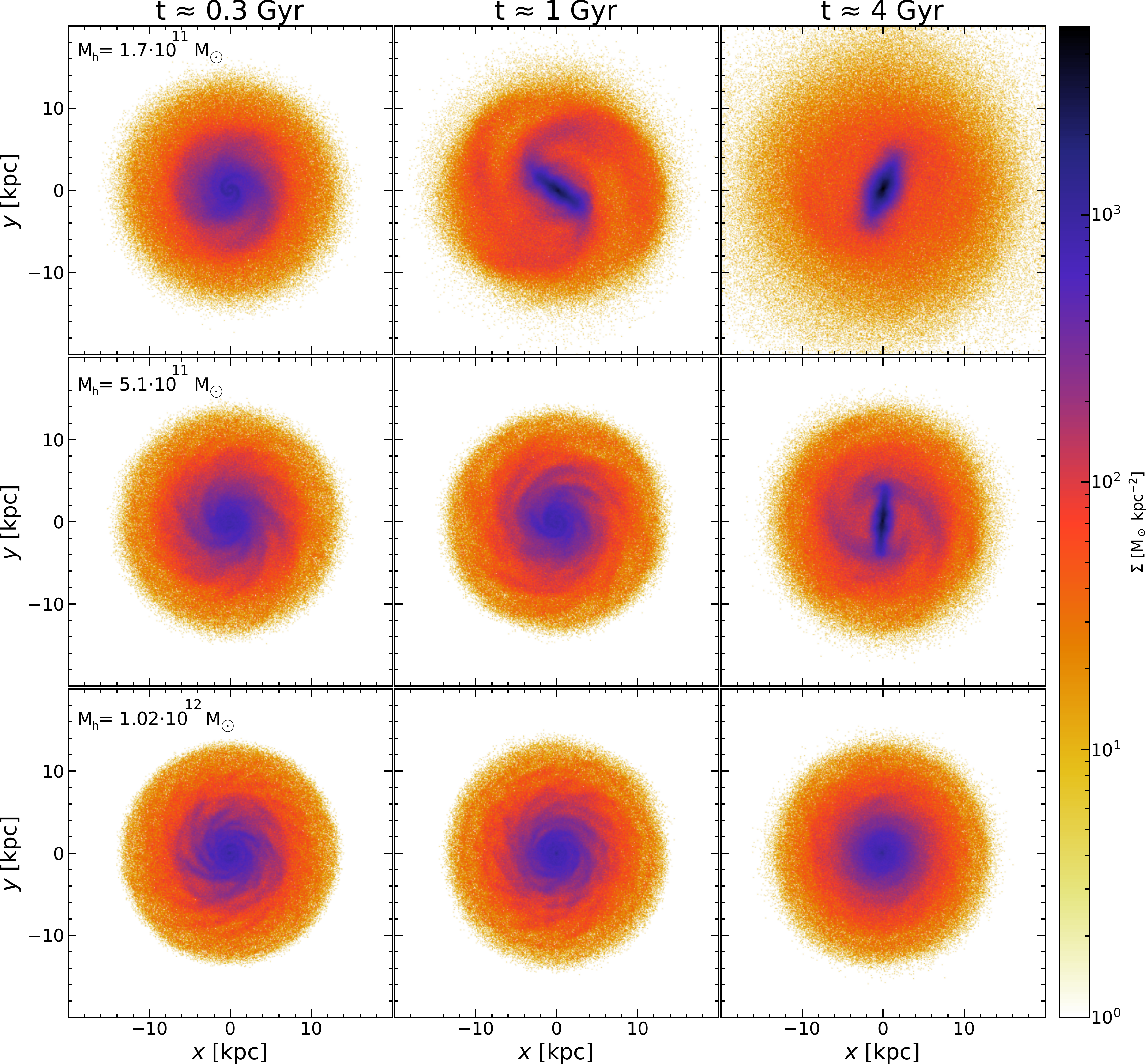}
    \caption{Face-on view of three of the different simulated galaxies, these are the disc corresponding to dark matter halo masses of  $1.7\times 10^{11}$ M$_\odot$,  $5.1\times 10^{11}$ M$_\odot$, and  $1.02\times 10^{12}$ M$_\odot$ in descending order along the rows. The columns show snapshots taken at $t\approx0.3$ Gyr, $t\approx1$ Gyr, and $t\approx4$ Gyr respectively, as indicated at the top, and the color shows the stellar density. In general we find that the number of spiral arms increases with the halo mass while their strength appears to decline.}
    \label{fig:evol}
\end{figure*}
\begin{figure*}
	\includegraphics[width=0.98\textwidth]{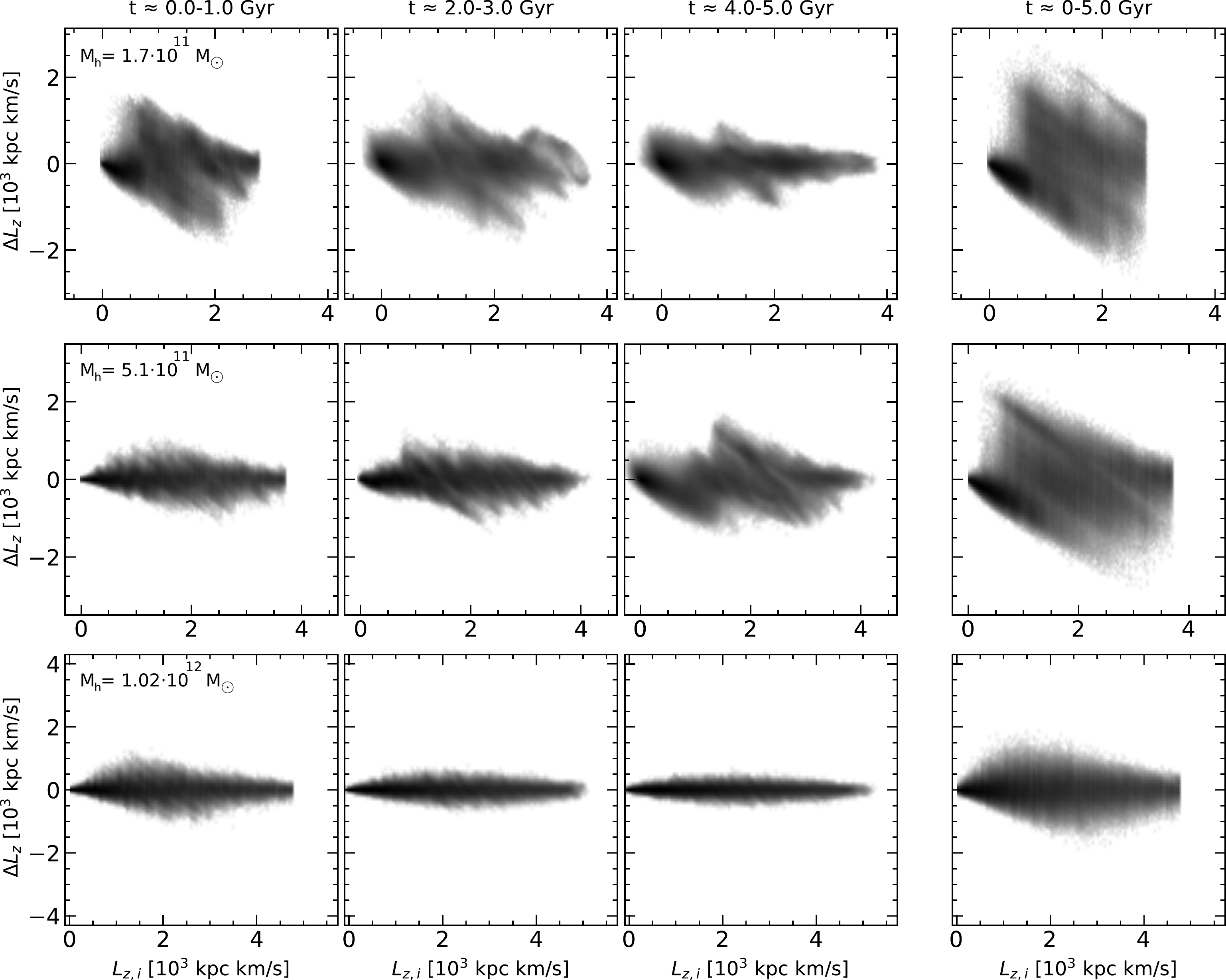}
    \caption{The change in angular momentum $\Delta L_z$ against initial angular momentum, $L_{z,i}$ compared at various times in the simulation. The simulations are the same as in Fig. \ref{fig:evol} and shown in the same order. The snapshot times compared are seen above the first row. The shading corresponds to number density. We can see that the weaker spirals seen in Fig. \ref{fig:evol} here generates less radial migration as is to be expected.}
    \label{fig:dlzlz}
\end{figure*}
\begin{figure*}
	\includegraphics[width=1\textwidth]{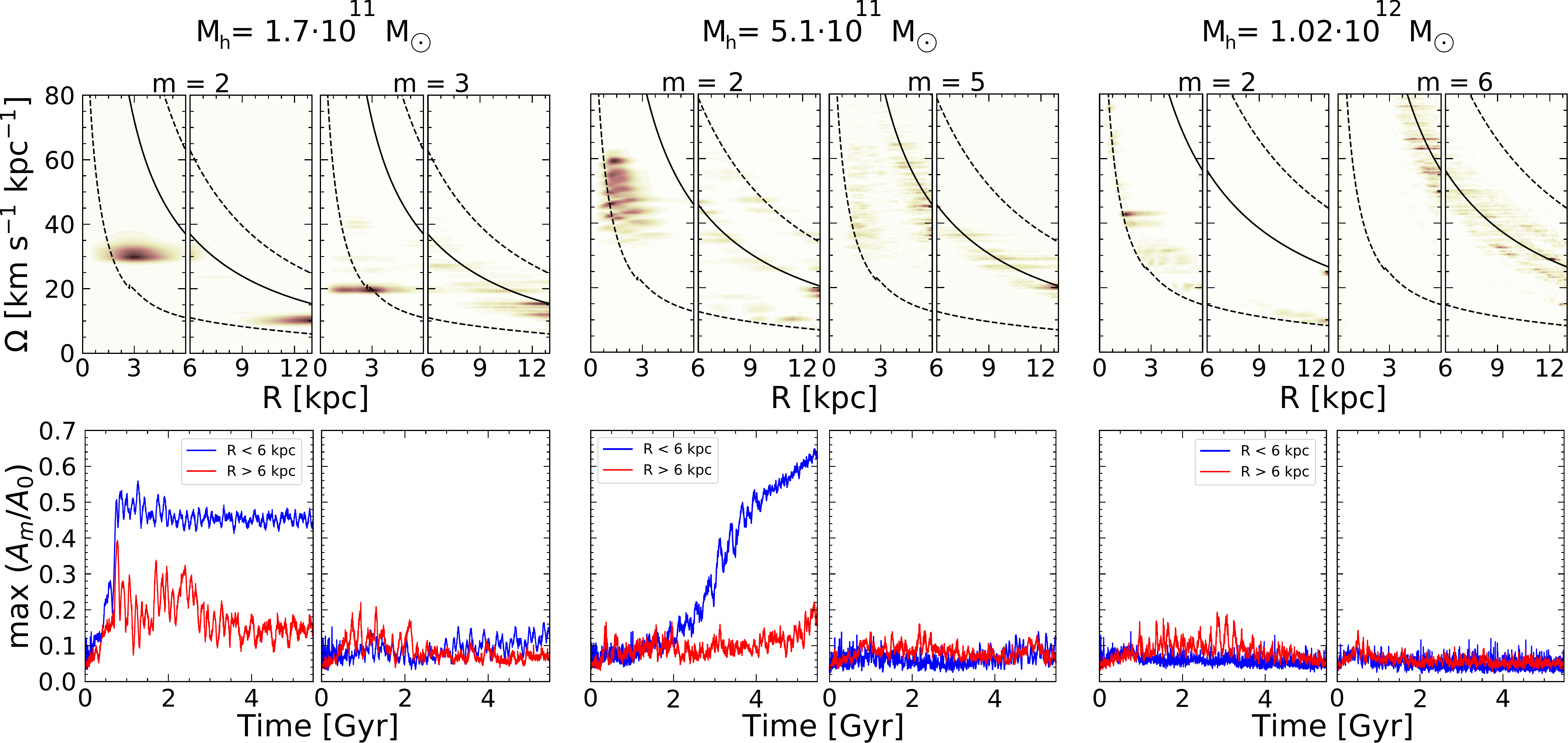}
    \caption{\textit{Upper row:} For each simulation in Fig.  \ref{fig:evol}, with mass indicated at the top, the power spectrum of the mode indicated above the plot. Shown are the angular velocities against radius. Solid lines indicate the co-rotation resonance with the disc and dashed lines mark the inner and outer Lindblad resonances. Multiple patterns can be seen across the various power spectra as horizontal lines since bars and spiral patterns have a fixed angular velocity. A bar can be seen in the two left-most simulations for $m=2$ at $\Omega \approx $33 km s$^{-1}$ kpc$^{-1}$ and $\Omega \approx $50 km s$^{-1}$ kpc$^{-1}$ for the left-most and middle simulations respectively. \textit{Bottom row:} The corresponding amplitudes of the modes. Just like the power spectra above it is divided into an inner and outer part divided at 6 kpc. The growth and decay of many different modes is visible. The two left-most simulations show the rise of a prominent bar in the $m=2$ mode. They are even more apparent within 6 kpc, reinforcing that it matches the patterns in the upper row figures. Some patterns are seen outside 6 kpc for these simulations as well and the lightest simulation shows a rise in an $m=3$ mode that dies after 2 Gyr. For the heaviest simulation, a very brief spike in $m=6$ can be observed in the amplitude but is difficult to discern in the spectogram.}
    \label{fig:fourier}
\end{figure*}
\begin{figure}
    \centering
	\includegraphics[width=0.4\textwidth]{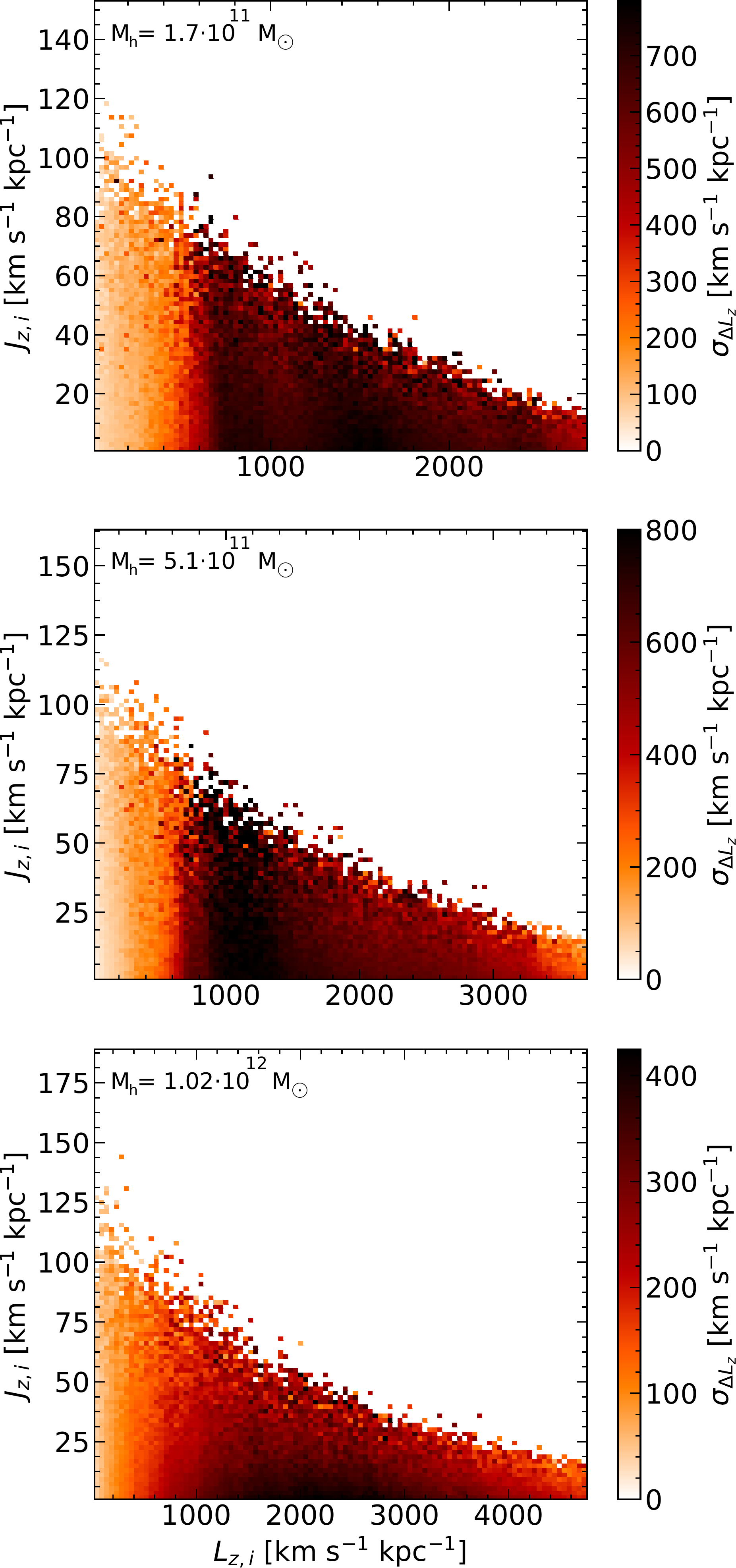}
    \caption{Initial vertical action and angular momentum of the same three galaxies shown in Fig. \ref{fig:evol} and in the same order. The space is binned 100x100 to show the standard deviation of angular momentum changes between $t = 0$ Gyr and $t \approx 5$ Gyr. For the top plot with lightest halo the migration appears to stay the same across all $J_z$ in contrast to the bottom plot with the heaviest halo which shows a clear decrease in radial migration as $J_z$ increases.}
    \label{fig:jzlz}
\end{figure}
\begin{figure*}
	\includegraphics[width=0.98\textwidth]{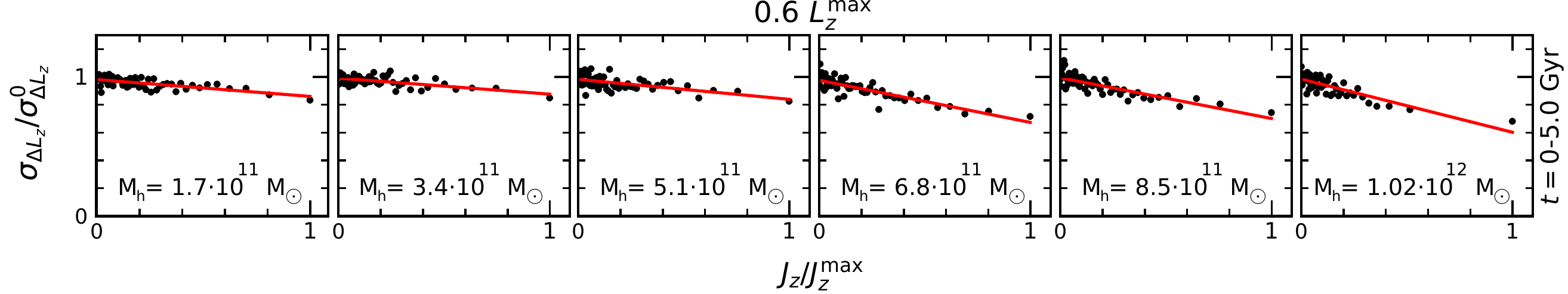}
    \caption{Standard deviation of the change in angular momentum, $\sigma_{\Delta L_z}$ normalized by the value at $J_z = 0$ against the vertical action normalized to the largest vertical action within the slice. Slices are vertical in Fig. \ref{fig:jzlz} taken at 0.6 of $L_z^\mathrm{max}$ and shown for a representative range of different halo mass simulations, ranging from the lightest on the left to the heaviest on the right. These slopes are taken when comparing angular momentum at $t\approx 0$ Gyr and $t\approx 5$ Gyr. The red lines show the linear fits from which we get the slopes.}
    \label{fig:slopes2}
\end{figure*}
\begin{figure*}
	\includegraphics[width=0.78\textwidth]{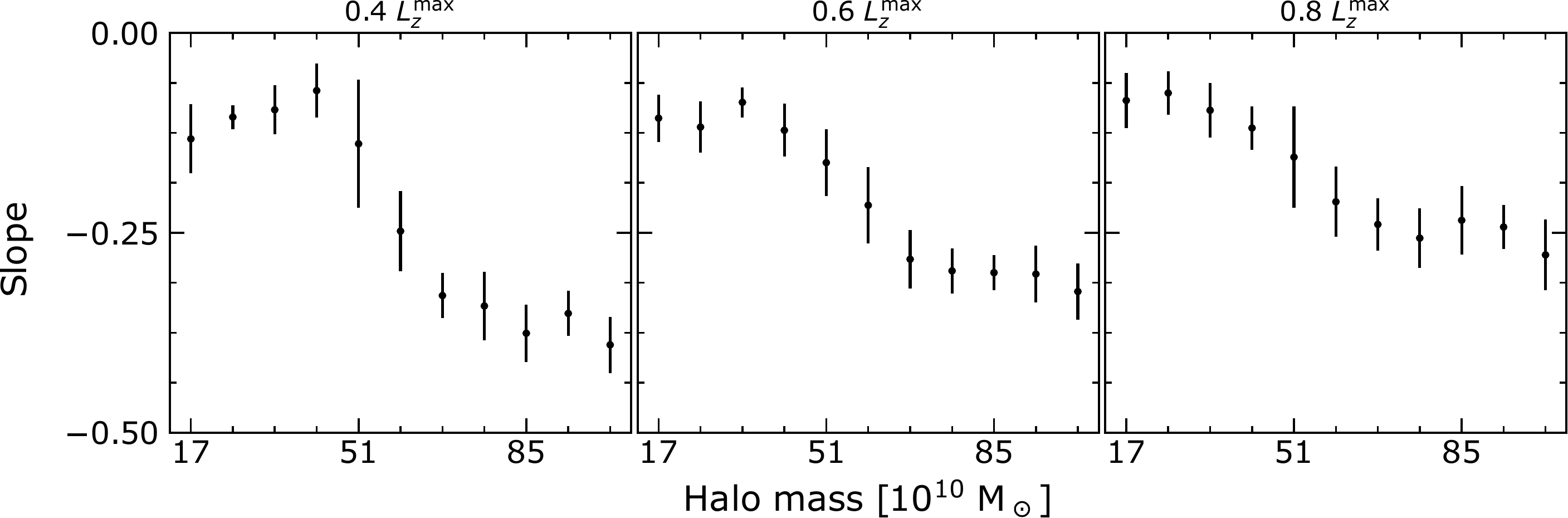}
    \caption{Compilation of all slopes calculated as described in Section \ref{sec:vertical} and seen as red lines in Fig \ref{fig:slopes2}. Slopes are plotted against halo mass showing a tendency for flatter gradients for disc-dominated systems and a steeper once for halo-dominated ones.}
    \label{fig:slopes}
\end{figure*}
\begin{figure*}
	\includegraphics[width=0.78\textwidth]{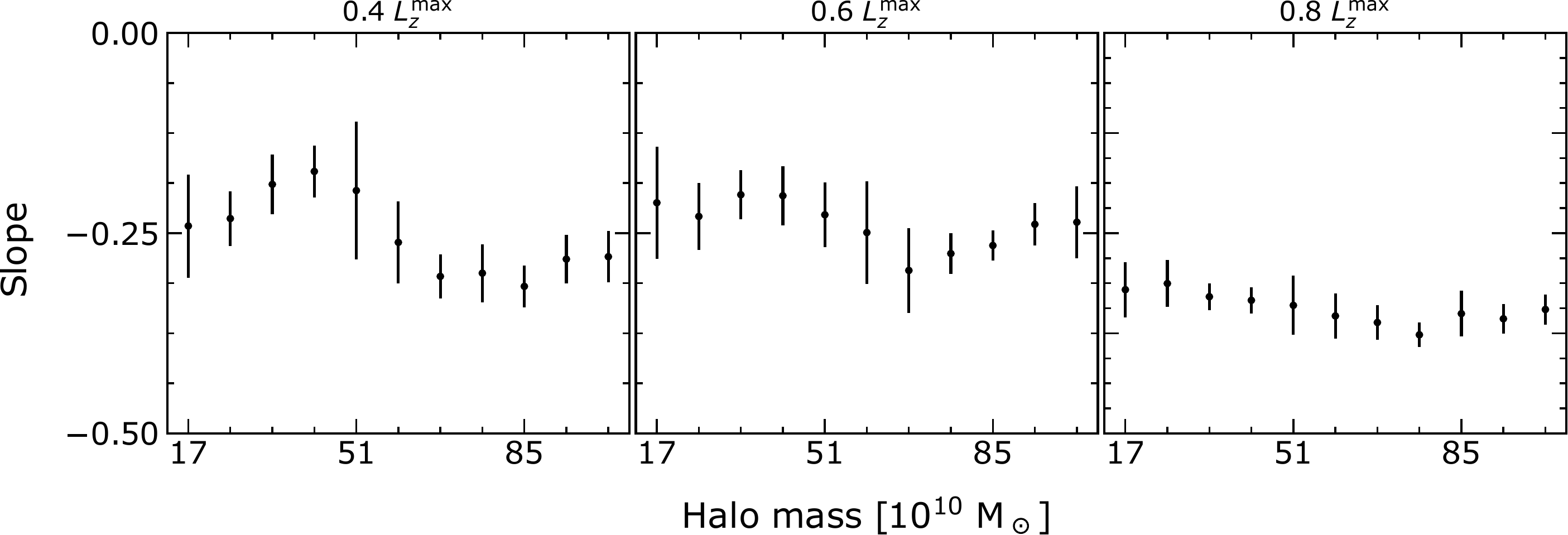}
    \caption{Similar to Fig. \ref{fig:slopes} using radial action instead of vertical action to investigate the gradient of migration. The radial action slope appears to be independent of halo mass, contrary to the vertical action counterpart.}
    \label{fig:slopesJR}
\end{figure*}

In order to create systems of varying number and strength of spiral modes, we generate galaxies where the dominance of the disc in relation to the halo is varied. This is achieved by simply increasing or decreasing the mass of the dark matter halo without changing any other parameters. The ratio of the radial force contributed by disc and dark matter halo at $z=0$, as a function of radius can be seen in Fig. \ref{fig:frate}. Halo masses are picked in order to have a large spread in this ratio, including both disc dominant systems and halo dominant ones. Fig. \ref{fig:frate} shows these ratios for the different halo mass simulations performed. The lightest halo starts at $1.7\times 10^{11}$ M$_\odot$ and increases by $0.85\times 10^{11}$ M$_\odot$ up until $1.02\times 10^{12}$ M$_\odot$. In accordance with literature \citep[e.g. ][]{dongia} the systems can be expected to form a larger number of spiral arms as the mass is increased. The strength of the spirals are expected to decrease as their number increases. For each system we generate 10 additional initial conditions varying only in random seed to give a sense of how the stochasticity of the initial conditions affects our results. In total this results in 121 simulations with 11 different seeds per halo mass used.

\subsection{Spiral strength analysis}\label{sec:fourier}
A key question is the link between how dominant the disc of a system is, the strength of corresponding bars/spirals, and the migration that ensues. To answer this the strength of the resonances that arise in the simulations must be measured. This is done using an extended Fourier analysis borrowing elements from \cite{press} and described thoroughly in \cite{roskar2012}. 

The disc is divided into annuli of equal widths such that the distribution of particles can be expanded into a Fourier series
\begin{equation}
    \Sigma(R,\phi) = \sum_{m=0}^\infty c_m(R)\exp(-im\phi_m(R)),
\end{equation}
with pattern multiplicity, $m$ and $\phi_m$ the phase of the $m$-th mode at the given radius. The coefficient $c_m(R)$ is given by
\begin{equation}
    c_m(R) = \frac{1}{M(R)}\sum_{j=1}^N m_j \exp(im\phi_j),
\end{equation}
with the mass in the annuli $M(R)$, the mass of particle $j$ as $m_j$, and the azimuth angle relative to the x-axis of the particle as $\phi_j$. $N$ is the total number of particles in the annuli. This method can be expected to identify bars at close radii and spirals at larger ones. Other azimuthal structures that can be identified through this method are ignored as any strong patterns appearing should be due to spirals or a bar. 

At this point only patterns at a given time can be detected but performing the above analysis at different points in time creates a time series out of the coefficients, $c_m(R)$, from which one can obtain a discrete Fourier transform
\begin{equation}
    C_{k,m}(R) = \sum_{j=0}^{S-1} c_j(R,m)w_j \exp(2\pi ijk/s),
\end{equation}
where $c_j(R,M) = c_m(R)$ at a given time $t = t_0 + j\Delta t$. The number of snapshots identified is the sample size, $S$, $C_{k,m}$ are the Fourier coefficients for the discrete frequencies $\Omega_k$, and $w$ is a Gaussian window function,
\begin{equation}
    w_j(x) = \exp(-(x-S/2)^2 / (S/4)^2),
\end{equation}
where $x$ is the current snapshot.

The frequency sampling is determined by the sample size, $S$, and the time between samples, $\Delta t$ such that
\begin{equation}
    \Omega_k = 2\pi \frac{k}{S\Delta t}m, \qquad k = 0,1,\dotsc,\frac{S}{2},
\end{equation}
avoiding high-frequency spectral leakage. Care is taken not to overstep the Nyquist frequency since $\Omega_\mathrm{Ny} = \Omega_{S/2}$. The complete power spectrum is then
\begin{equation}
    P(\Omega_k, R) = \frac{1}{W}\left[|C_k(R)|^2 + |C_{S-k}(R)|^2 \right], \qquad k = 1,2,\dots,\frac{S}{2}-1.
\end{equation}
The power is normalized by $W = \sum_{j=0}^S w_j$. Using $P$, $\Omega_k$, and $R$ it is possible to construct a contour plot of the power spectrum, allowing identification of the pattern speed and extent of a certain mode. 

In addition to this, the strength of a certain mode over time can be retrieved through the absolute value of the coefficients, $c_m(R)$, at a given time. This gives the amplitude of the wave,
\begin{equation}
    A_m(R) = |c_m(R)|.
\end{equation}

Dividing the maximum value of the amplitude by the amplitude of the zeroth mode for each snapshot gives the growth and evolution of a certain mode. However, this will only identify the strongest pattern for a given mode at any given time. That is, if for example the bar forms it will overshadow a previously identified $m=2$ spiral. For this reason, the disc is separated into an inner, $R < 6$ kpc, and outer $R \geq 6$ kpc region. Generally, the bar will be the dominant feature in the inner region and will not appear in the outer one, leaving spirals to be identified. 

\section{Results}\label{sec:results}
\subsection{Disk evolution and migration}
The results are shown at different times up to $\sim$5 Gyr, where it is ensured that non-axisymmetric interactions have taken place. That is, spiral arms have grown, churned, and faded, leaving a cumulative radial migration effect upon the galaxy. Since our simulations are pure $N$-body, there is no need to integrate further as new spiral arms are not excited at later times. Three different simulations are presented in Fig. \ref{fig:evol} and show the evolution of secular resonances. These three galaxies are chosen to show examples of disc dominance, halo dominance, and something intermediate. It therefore shows different number and strengths of spiral arms, as we expect. It is clearly visible that as the dark matter halo starts to dominate, the disc is unable to form larger, grand-design type spiral arms and instead form many weaker spiral arms. When the disc dominates there is also a significant radial `puffing' up of the disc leading to a radially extended structure at later times.

The evolution is shown at three different points in time. The first two snapshots are taken from within the first fifth of the integration time ($t\approx 0.3$ Gyr and $t\approx1$ Gyr respectively) as there is very little significant secular evolution beyond that point. The last snapshot is at $t\approx4$ Gyr, and does not change significantly after. 

To allow for secular evolution and its effect on the migration, a few different combinations of `initial' and `final' times are investigated in $\Delta L_z$-$L_{z,i}$ space, that is the change in angular momentum compared to the initial angular momentum. This can be seen in Fig. \ref{fig:dlzlz} and here spirals arms manifest as diagonal lines near the co-rotation radius of a spiral or bar. Particles located exactly at co-rotation show no change to the angular momentum and those inside co-rotation would migrate outwards and vice-versa. So particles inside have a positive change in angular momentum and stars outside have a negative one, giving rise to the diagonal ridges.

Fig. \ref{fig:dlzlz} shows that as the simulations produce different number and strength of spiral arms, the radial migration that takes place also differs. It is clear that a more massive halo produces many smaller spiral arms, as seen in Fig. \ref{fig:evol}, which produce the weak diagonal features. When the halo dominance grows there is much less total migration as evident in the spread in $\Delta L_z$.  This result is to be expected since a weaker arm produces a smaller torque and hence, smaller changes to the angular momentum.

As can be seen in the upper two rows of Fig \ref{fig:evol}, a bar eventually forms in some of these simulations and appears to do so whenever the galaxy can readily form spirals. The prominent formation of a bar could potentially be prevented with a stronger bulge, creating an inner Lindblad resonance to serve as a barrier against formation. Once the bar forms it grows to be a rather large $m=2$ resonance which has a tendency to overshadow smaller spiral features. 

We can investigate the appearance of a bar and spirals more closely using the fourier analysis outlined in Section \ref{sec:fourier}. The power spectrum and amplitudes of the dominant modes are shown in Fig. \ref{fig:fourier}. The growth of a stable bar is seen very clearly in the amplitude of the $m=2$ mode in the lightest two halo mass simulations. The evolution of the $m=2$ mode is shown for each simulation, as this is usually the dominant mode when a bar forms. Also shown is the evolution of other prominent modes in each respective simulation. The lower mass simulation has a few occasions with $m=3$ modes which are rather short-lived. The noisiness observed in this plot could well be due to the presence of material arms which are short-lived and therefore are difficult to capture in the power spectrum, but are very readily visible in plots such as Fig. \ref{fig:evol} and Fig. \ref{fig:dlzlz}. It is still possible to observe a hint of a large $m=6$ mode at the start of the heaviest halo mass simulation, which is in line with the predictions for the multitude of spiral arms when the disc dominance is low.


\subsection{Migration and action}\label{sec:action}
\subsubsection{Vertical action}\label{sec:vertical}
The appearance of different spiral structures depending on dominance of the disc has been established in our simulations. In Section \ref{sec:theory} we made the argument that the amount of radial migration at various heights above the midplane of the disc could be linked to the spiral strength. To characterise the vertical distribution of our migrators we use vertical action instead of position or velocity as the vertical action does not oscillation on orbital timescales while position and velocity does. The data is binned in initial vertical action, $J_z$, and initial angular momentum, $L_z$, as a proxy for radius. For each bin we calculate the standard deviation of the change in angular momentum, $\sigma_{\Delta L_z}$, which quantifies the amount of radial migration or radial migration efficiency. The result of this is seen in Fig. \ref{fig:jzlz}. For the lightest halo which showed strong spirals and an eventual bar, there is strong migration at almost all $J_z$ and in almost all of the disc in $L_z$, save for the innermost parts. The migration at high vertical action gets weaker as the halo grows in mass and once the halo becomes relatively dominant in the bottom plot radial migration appears to decrease as the vertical action is increased. 

The different behaviours described here can be quantified more clearly by recognising that the discussion of migration at various vertical actions is a discussion of a slope in $\sigma_{\Delta L_z}$ with $J_z$. To investigate any possible radial dependencies three slices in $L_z$ with widths (1/50)$L_z^\mathrm{max}$ at 0.4, 0.6, and 0.8 of the maximum $L_z$ are taken, while the innermost cut at 0.2 is omitted as it contains orbits which belong to a bar once it forms. The disc does not extend beyond $\sim$15 kpc, so the maximum $L_z$ corresponds to that of near circular orbits at this radius. To clarify, we take the particles within a (1/50)$L_z^\mathrm{max}$ width of the specified locations regardless of their $J_z$, creating three ``slices". The separation into these three different slices allows for the comparison of migration as a function of $J_z$ at different distances from the centre of the galaxy. The particles in the slices are binned again in the $J_z$ direction, with 300 particles in each bin. We perform a least squares linear fit for $\sigma_{\Delta L_z}$ as a function of $J_z$ where the gradient of the line we fit will reflect how the vertical action of a particle affects its radial migration.

However comparing the slope of the lines in $\sigma_{\Delta L_z}$ and $J_z$ as is would not prove very informative, because different parts of a galaxy in $L_z$ reach different maximum $J_z$ due to the difference in the gravitational potential, as clearly seen in Fig. \ref{fig:jzlz}. In order to correct for this we divide $J_z$ in the slices by close to their largest value\footnote{The normalizing value is the median of $J_z$ values in their second highest bin in the slice, which is chosen to avoid outliers. Our results are robust to the specific choice of normalization. Normalizing against a different percentile would not affect the result as it is the behaviour across the disc which is of interest.}. This correction normalizes $J_z$ across the disc.

It is also problematic that the different simulations have different ranges of $\sigma_{\Delta L_z}$, since total migration decreases when the spiral arms become weaker (i.e., with increasing halo mass). For this correction $\sigma_{\Delta L_z}$ is divided by its mean around $J_z = 0$ in each slice, which normalizes the results across different dark matter halo masses.

We wish to know the vertical gradient of radial migration regardless of the total amount of radial migration or galactocentric radius, which this normalization will allow. 

To give an example of this normalization Fig. \ref{fig:slopes2} shows the slices at 0.6 $L_z^{\rm max}$. The slices and linear fits are seen for simulations with six different dark matter halo masses, ranging from lightest to heaviest. The gradient in radial migration as a function of vertical action is shown as red lines. As the halo mass is increased and the disc dominance is decreased, we see that the slope becomes steeper and migration less significant for high $J_z$ particles, making the changes hinted at in Fig. \ref{fig:jzlz} clear.


The three slices of the disc are compiled to show the slopes in Fig. \ref{fig:slopes} for the various dark matter halo masses and at the different angular momenta. The slope is calculated for each random initialisation with the same halo mass setup and the standard deviation of those values is used to give the error bars shown. 

Fig. \ref{fig:slopes} shows a quantified version of the arguments made above regarding the vertical gradient of the radial migration. For the lower halo masses, we have strongly self-gravitating discs that result in normalized slopes closer to zero. These cases of low halo mass correspond to the situation seen in the top plot, and to some extent the middle, of Fig. \ref{fig:jzlz}. A separate regime appears as the halo mass increases and the disc is less dominant. Now the slopes are larger, showing a radial migration bias for low-$J_{z,i}$ stars much like the `provenance bias' of \citet{vera-ciro14} discussed in Section \ref{sec:theory}. There is a smooth transition between the two different regimes discussed and the trend is too strong to be explained by the random scatter that we can see.


It can be clearly seen in Fig. \ref{fig:slopes} that by changing the relative gravitational influence of the halo and the disc in a galaxy simulation, and therefore the resulting spiral morphology, you can change the extent to which stars with lower vertical action are preferentially radially migrated.



\subsubsection{Radial action}
It is also useful to perform an investigation into the behaviour of radial migration as a function of the radial action in our simulations. Just as the vertical action is a good measure of how vertically heated an orbit is (see Section \ref{sec:theory-actions}) the radial action is a radial equivalent and is related to the eccentricity of a stellar orbit. \citet{kathryne} found that radial migration was less efficient in populations with larger radial velocity dispersion. A similar result was also found by \citet{solway} who looked at the angular momentum changes compared to initial eccentricity. They showed that angular momentum changes were larger for particles with smaller eccentricities. 

These results can be understood within the theory as well. If the eccentricity of a particle is large its angular velocity will oscillate as it orbits and will only match that of a steadily rotating spiral for a brief period of time making it less likely to enter co-rotation with the spiral. The more circular the orbit of the particle the more readily it responds to the resonance. This argument is similar to the the arguments made regarding the vertical bias of migrators with a key difference being that a vertical oscillation brings the particle away from the disc which a radial oscillation will not.

To perform this analysis we use the same procedure as in the preceding section for vertical action. Three slices at separate $L_{z,i}$ are cut in the space of $J_{r,i}$ and $L_{z,i}$. Within these slices the standard deviation of the angular momentum changes, $\sigma_{\Delta L_z}$, are taken within bins of 300 particles each. The linear slope is normalized in the same manner and calculated for each halo mass simulation and for all the different random initializations. 

We present the result in Fig. \ref{fig:slopesJR}. Here, the gradient of migration shows no strong trend with increasing halo mass. The value of the slope is almost constant at around -0.25, which means that radial migration is more efficient for less eccentric orbits, in agreement with the results and theory stated in the previous paragraphs. The innermost part of the galaxy contains orbits that belong to the bar and has strongly nonlinear behaviour between the radial action and the migration efficiency. Fitting a line to that region of the disc yields a large scatter in slope values which give no real information on the migration process and as such is not shown here.

When comparing Fig. \ref{fig:slopesJR} with Fig. \ref{fig:slopes} it is clear that there is a difference in the response to disc dominance and stronger/weaker spirals between the two axes which could be caused by the former being confined to the disc while the other is not.

\subsubsection{Action conservation}
\begin{figure}
\vspace{5pt}
    \centering
	\includegraphics[width=0.47\textwidth]{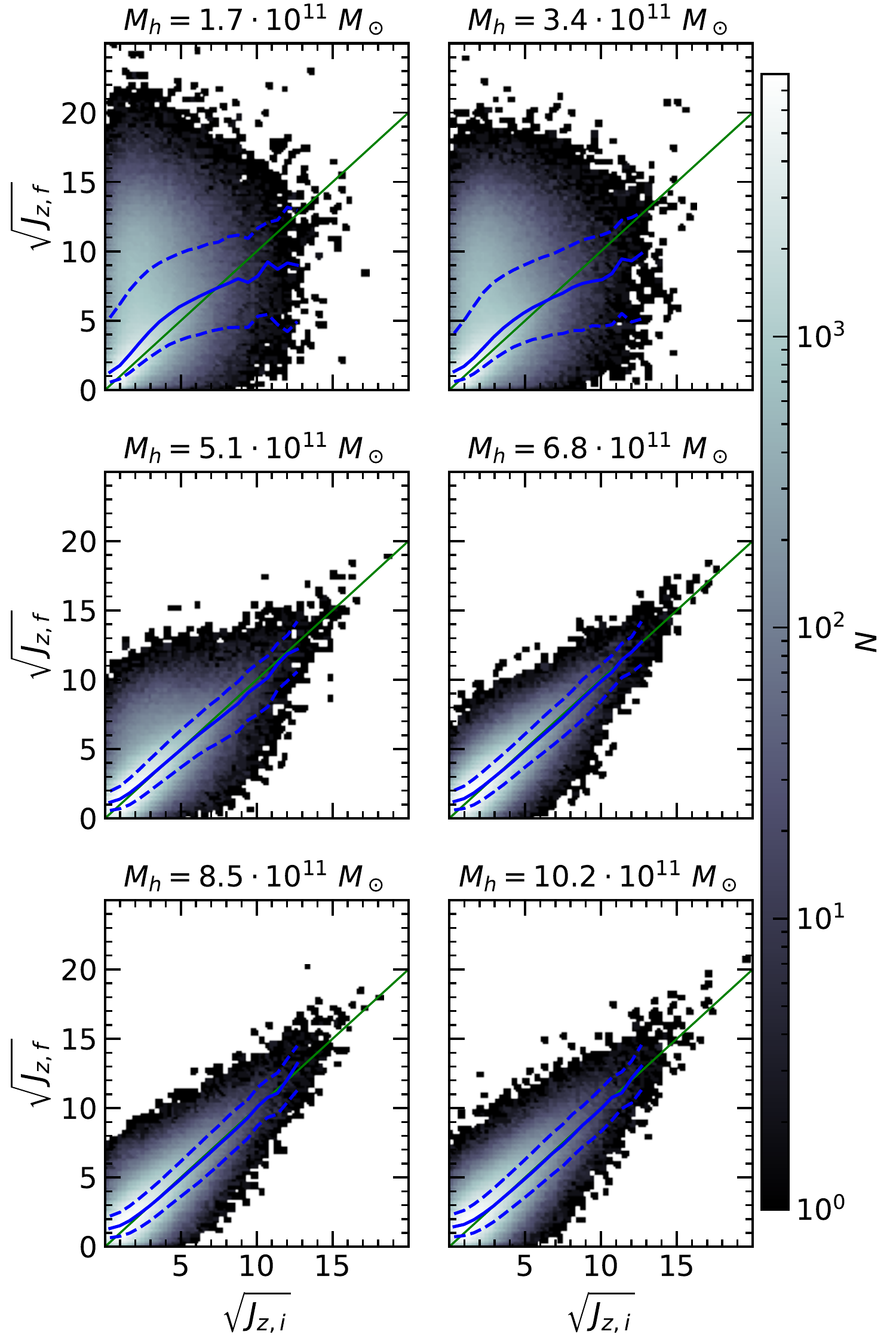}
    \caption{Number density distribution of $\sqrt{J_{z,i}}$ and $\sqrt{J_{z,f}}$ in units of (kpc km/s)$^{1/2}$ calculated at $t = 0$ and $t\approx 3$ Gyr respectively. Also shown is the median value of $\sqrt{J_{z,f}}$ in 20 bins of $\sqrt{J_{z,i}}$ as a solid blue line. Dashed blue lines show the 1$\sigma$ range and exact action conservation $J_{z,f} = J_{z,i}$ is shown in green. The results are shown for the same simulations as in Fig. \ref{fig:slopes2} and indicated above each individual plot.}
    \label{fig:jzden}
    \vspace{-15pt}
\end{figure}
\begin{figure}
\vspace{5pt}
    \centering
	\includegraphics[width=0.47\textwidth]{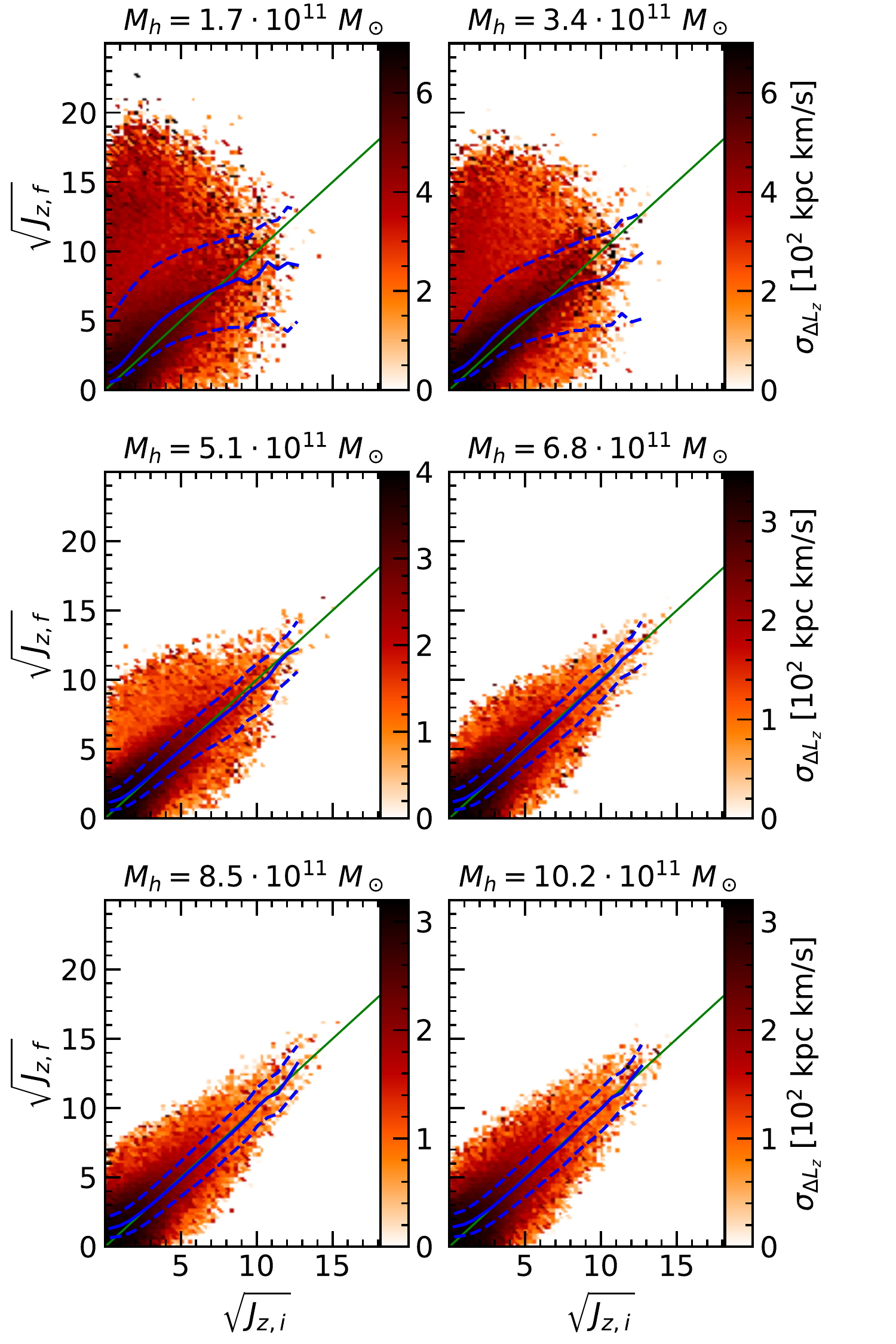}
    \caption{Same as Fig. \ref{fig:jzden} with the colour corresponding to the same radial migration efficiency as Fig. \ref{fig:jzlz} instead of number density.}
    \label{fig:jzrm}
    \vspace{-15pt}
\end{figure}
 \begin{figure*}
	\includegraphics[width=0.72\textwidth]{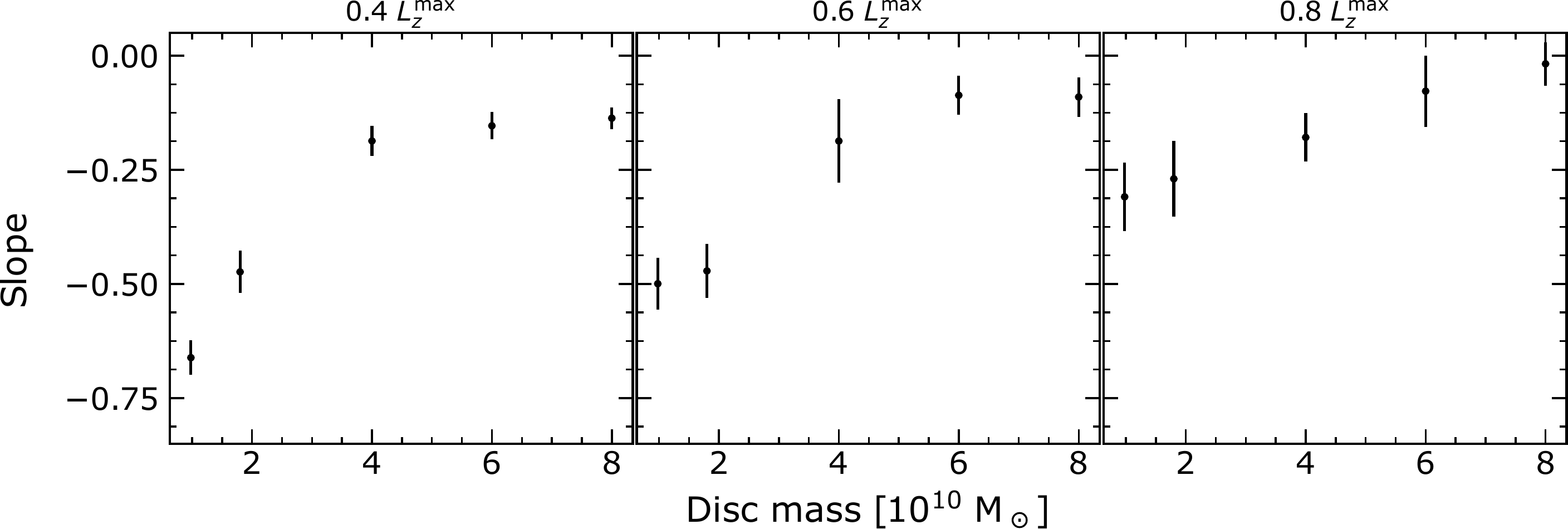}
    \caption{Similar to Fig. \ref{fig:slopes} but instead slopes are plotted against disc mass, meaning that the disc becomes more dominant along the $x$-axis. The simulations used to evaluate the slopes are recreations of the {\tt HD-MW} simulation of \citet{vera-ciro16}. It is clear that the slope flattens with disc dominance.}
    \label{fig:slopesVC}
    \vspace{-15pt}
\end{figure*}
In the previous sections the radial migration has been compared between snapshots taken at $t = 0$ and $t\approx 5$ Gyr, while the vertical action has been calculated for $t = 0$, assuming it has not changed by the time the particle migrates. To be certain of this we must investigate how well the vertical action is conserved over the duration. 
Conservation of vertical action was studied by \citet{solway} who concluded that vertical action is conserved on average while it may change for individual particles. This work was expanded on by \citet{vera-ciro_actions} who compared two $N$-body spiral galaxies. One was set up to form multi-armed spirals and the other, more Milky-Way like galaxy, to form a bar. They found that with the formation of a bar, the actions are not very well conserved.

For six of our different halo mass simulations we compare the value of the vertical action at $t\approx 0$ Gyr and $t\approx 3$ Gyr (called here $J_{z,i}$ and $J_{z,f}$ respectively)\footnote{As in e.g. \citet{trick} we use $\sqrt{J}$ to reveal structure at lower actions.} as a binned density map in Fig. \ref{fig:jzden}. Beyond $t\approx 3$ Gyr the data is only blurred by noise. We include the median and 1$\sigma$ range of $J_{z,f}$ for 20 bins in $J_{z,i}$ as blue solid and dashed lines respectively. This can be compared to exact action conservation $J_{z,f} = J_{z,i}$ shown as a solid green line. 

The lightest halo masses in Fig. \ref{fig:jzden} show that when the disc is dominant and a bar forms, as seen in Fig. \ref{fig:evol}, there is a large spread in vertical action around action conservation, in agreement with \citet{vera-ciro_actions}. However, the median lies close to the line $J_{z,i} = J_{z,f}$. The heavier halos show a smaller spread about the line of conservation, and in agreement with \citet{solway} their medians align almost exactly with $J_{z,i} = J_{z,f}$. While individual bins show a large spread the median is sufficiently close to $J_{z,i} = J_{z,f}$ that $J_{z,i}$ is a good predictor of a typical $J_{z}$ for all halo masses.

It is however clear that vertical action can, for individual particles, change significantly. We investigate to what extent radial migration can cause this by again comparing the same initial and final vertical actions in a binned histogram as above but coloured by radial migration efficiency, $\sigma_{\Delta L_z}$, instead of density. This is shown in Fig. \ref{fig:jzrm} with the median and 1$\sigma$ range from the density histogram overplotted. 

From the discussion in Section \ref{sec:theory}, churning is not expected change the radial action. Fig. \ref{fig:jzrm} shows that the most significant migration is around the line of conservation of $J_z$ in all simulations. For the lighter halos there is still a spread to larger $J_{z,f}$ where migration efficiency is also lower, so the most migrated particles do not change their action significantly. Thus the change of vertical action is likely due to heating in the bar or a similar process and not due to the migration itself. The fact that most of the particles are located near the line of conservation, and that this is where most of the radial migration occurs, supports the idea that the behaviour in Fig. \ref{fig:slopes} is a gradient for radial migration by churning and is not significantly polluted by another process

\subsection{Comparison to Vera-Ciro}
As was shown in Section \ref{sec:vertical} the vertical gradient of radial migration changes depending on the disc dominance of the initial conditions. Despite this, literature results generally claim that there is less migration at larger vertical excursions as discussed in Section \ref{sec:theory}. Particularly \citeauthor{vera-ciro14} (\citeyear{vera-ciro14}, \citeyear{vera-ciro16}) has studied the radial migration in simulations with different spiral morphologies where they find a significant negative vertical gradient for radial migration regardless of spiral morphology, contrasting our results, and we wish to understand our results within this context. We have shown that high vertical action will prevent a star from being churned in cases where the disc is less dominant in the galaxy. The galaxies that are set up in these papers may simply be in a regime where this is the case. In order to investigate this thoroughly we have chosen to recreate and compare with the galaxy from \citet{vera-ciro16} labelled {\tt HD-MW} or `Milky Way-like'. The dominance of the disc is changed directly through the disc mass and we have used five setups with $[0.245,0.45,1,1.5,2]$ times the original disc mass, $4\times 10^{10}$ M$_\odot$. Ten extra seeds are again generated for each simulation to test for stochastic robustness and the same procedure described in Section \ref{sec:vertical} is performed to generate Fig. \ref{fig:slopesVC}. Now the slope flattens along the $x$-axis as they are plotted against disc mass instead of halo mass. It can be seen that while there is a tendency for larger vertical action to reduce radial migration at $M_d = 4\times 10^{10}$ M$_\odot$, it is not very prominent and becomes stronger/weaker if the mass of the disc is smaller/larger. 

\begin{figure}
\vspace{5pt}
    \centering
	\includegraphics[width=0.4\textwidth]{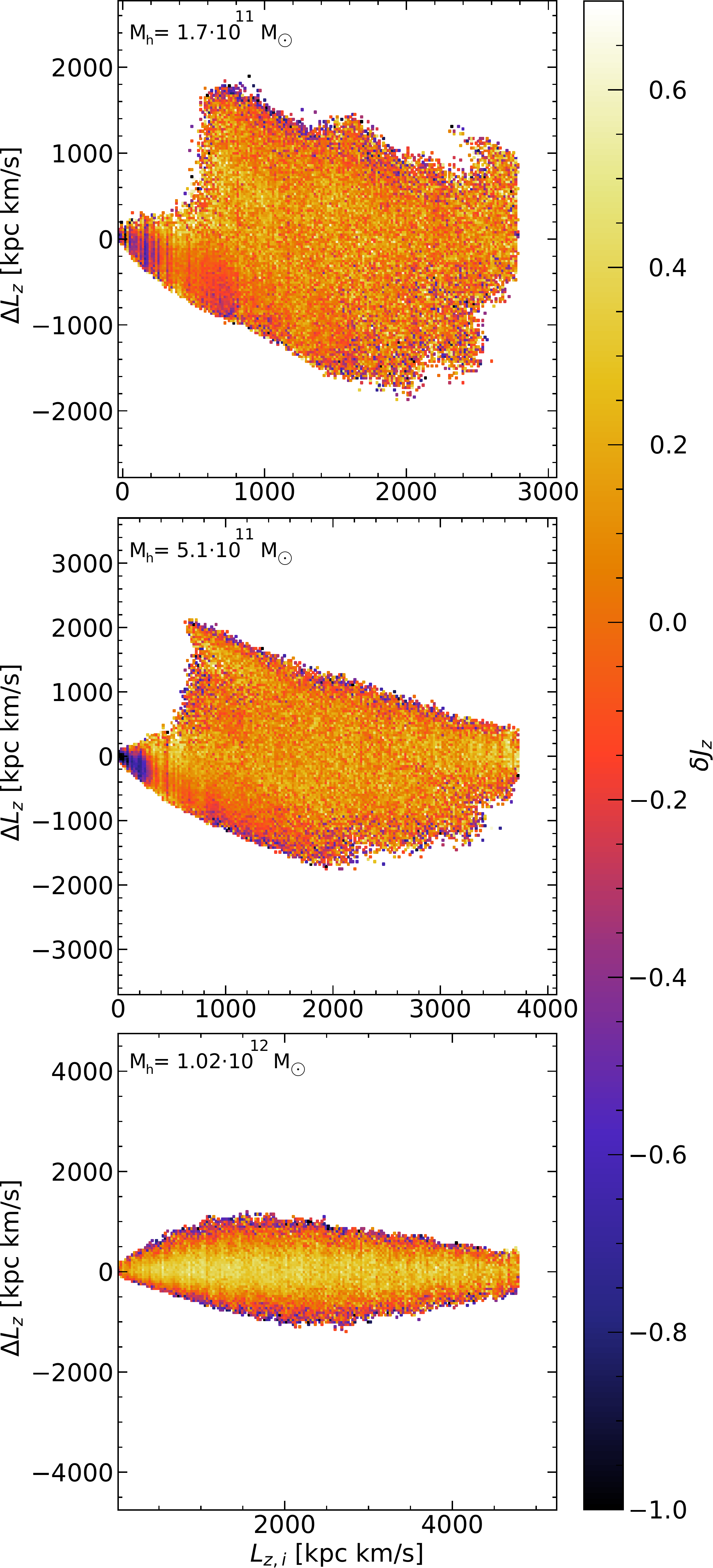}
	\vspace{-3pt}
    \caption{Same as the rightmost column of Fig \ref{fig:dlzlz} but coloured by the log of the vertical action divided by the mean of the vertical action at that angular momentum. In the more disc dominated systems we can see that the particles that migrate do not have very significantly different $J_z$ than those that do not migrate as much. When the disc becomes less dominant, migration is more prominent among particles of low vertical action.}
    \label{fig:vcplot}
    \vspace{-15pt}
\end{figure}

In Fig. 4 of \citet{vera-ciro16} the distribution of velocity dispersion is shown in terms of initial guiding center radius, $R_{g,i}$, and fractional change in guiding center radius, $\delta R_g = \ln (R_g/R_{g,i})$ between an initial time and at $t \sim 2$ Gyr. The velocity dispersion is the initial one in units of the average velocity dispersion at that radius. That is 
\begin{equation}
    \delta\sigma_{z,i} = \ln\ \frac{\sigma_{z,i}(R_{g,i},\delta R_g)}{|\sigma_{z,i}(R_{g,i})|}.
\end{equation}
Using this they claim a lower velocity dispersion for particles of larger $\delta R_g$. For comparison the three simulations shown in Fig \ref{fig:dlzlz} are investigated to show the distribution in vertical action, using a similar procedure such that
\begin{equation}
    \delta J_z = \ln\ \frac{J_{z,i}(L_{z,i},\Delta L_z)}{|J_{z,i}(L_{z,i})|}.
\end{equation}
This equation normalizes the vertical action in a manner similar to how the normalization in $J_z$ was carried out for the slope calculation. The result of this can be seen in Fig. \ref{fig:vcplot}. It is clear also here that disc dominance flattens the gradient in radial migration and vertical action, as seen in the previous results. 




\section{Conclusions}\label{sec:conc}
In this paper we have studied radial migration efficiency as a function of vertical and radial action as well as how these functions depend on disc dominance. We have used a large suite of $N$-body simulations of isolated galaxies where we have varied the total amount of mass within the dark matter halo in order to change the dominance of the disc and thereby create spiral structures of different number and strength. 

The main focus of this study has been to identify a relationship between disc dominance and the subsequent radial migration for different heights above the disc midplane, measured through the vertical action, $J_z$. This link has not been previously established and we have shown that if the disc of a galaxy is made more dominant, and therefore the spiral arms within it made stronger and fewer, radial migration can occur at larger vertical extents. This adds to our current understanding of radial migration in galaxies like the Milky Way and is instrumental in the implementation of radial migration in analytical studies of galactic dynamics.

Previous studies that have looked at the vertical gradient of radial migration have not reached the same conclusion \citep{solway,vera-ciro14,halle2015,vera-ciro16}. To understand this we have studied the results of \citet{vera-ciro16} extensively within the context of our findings as they also investigate the role of different non-axisymmetric patterns on radial migration, partly in terms of vertical velocity dispersion, $\sigma_z$. They find that migrators are a subset of stars with small vertical velocity dispersions regardless of the observed spiral morphology achieved, in contrast to our findings. One of their simulations was reconstructed here and tested with different disc dominance to produce the results we have seen. We show that we are able to create different vertical gradients of radial migration depending on the disc dominance in this case as well. Within these findings the results of \citet{vera-ciro16} likely stems from a sampling of a specific part of the parameter space where there is less variation in the vertical gradient of migration.

In addition we studied the gradient of radial migration with radial action as a proxy for eccentricity. This relationship has previously been studied by \citet{kathryne} who showed that radial migration was less efficient in populations with larger radial velocity dispersion and before that by \citet{solway} who compared $\Delta L_z$ with eccentricity to find similar results. Our simulations are in agreement with these two results regardless of the initial conditions. The strength of the spiral arms does not have a noticeable effect on which particles are radially migrated. 

Our result for the radial action and vertical action contrast one another. This means that there is a difference in the response to co-rotation with increasing action due to the direction being along the disc or orthogonal to it. This discrepancy is interesting and invites further analysis.

The vertical action is less well conserved in our disc dominant simulations with a bar. The method by which the vertical gradient in Fig. \ref{fig:slopes} is determined relies on the initial vertical action, $J_{z,i}$, reflecting the vertical action around the time of migration. However, as the action is well conserved on average for the majority of the simulations and only slightly deviates from average conservation in our most disc dominant simulations, our vertical gradient determinations should not be severely affected. One method to eliminate this issue would be to determine the vertical action as closely as possible to the time of migration, a feat which demands more detailed analysis.

These findings matter for stars with large vertical excursions, which are typically part of the oldest populations in the disc. Our results imply that, if the disc is sufficiently dominant, stars can be migrated despite being part of such a vertically extended population. Any interpretation of the distribution of stars as a function of age in the Milky Way would be affected by this, and it has significant implications for the formation and evolution of thick discs.

Previous work studying radial migration and its implications have made use of various combinations of analytical models and simulations both hydrodynamical and $N$-body, full or using static potentials. Here we present the combined result of over a hundred full $N$-body simulations to reduce the stochasticity of our findings. Results like these which aid in describing the nature of radial migration are necessary to further the analytical modelling which relies on descriptions provided by studies of this kind. 

Radial migration is still not fully understood and will certainly be the subject of future studies both numerical and analytical. Through the results we have presented here, the desired vertical gradient of radial migration can be tuned through the choice of relative gravitational strength of the disc to that of the dark matter halo.

\section*{Acknowledgements}
We thank members of Lund Observatory as well as an anonymous referee for helpful comments and ideas. Computations for this study were performed on equipment funded by a grant from the Royal Physiographic Society in Lund. PM is supported by a research project grant from the Swedish Research Council (Vetenskapr{\aa}det). DH and PM gratefully acknowledge support from the Swedish National Space Agency (SNSA Dnr 74/14 and SNSA Dnr 64/17).





\bibliographystyle{mnras}
\bibliography{references}







\bsp	
\label{lastpage}
\end{document}